# Proposal to Search for Magnetically Charged Particles with Magnetic Charge 1e*

Michael K. Sullivan[1] and David Fryberger[1]

[1]SLAC National Accelerator Laboratory
2575 Sand Hill Rd. MS-51 and MS-39, Menlo Park, CA, USA 94025
email address: sullivan@slac.stanford.edu
email address: fryberger@slac.stanford.edu

A model for composite elementary Standard Model (SM) particles based upon magnetically bound vorton pairs, briefly introduced here, predicts the existence of a complete family of magnetically charged particles, as well as their neutral isotopic partners (all counterparts to the SM elementary particles), in which the lowest mass (charged) particle would be an electrically neutral stable lepton, but which carries a magnetic charge equivalent to 1e. This new particle, which we call a magneticon (a counterpart to the electron) would be pair produced at all $e^+e^-$ colliders at an $E_{cm}$ above twice its mass. In addition, PP and PPbar colliders should also be able to produce these new particles through the Drell-Yan process. To our knowledge, no monopole search experiment has been sensitive to such a low-charged magnetic monopole above a particle mass of about 5 GeV/c². Hence, we propose that a search for such a stable particle of magnetic charge 1e should be undertaken. We have taken the ATLAS detector at the LHC as an example in which this search might be done. To this end, we have modeled the magnetic fields and muon trigger chambers of this detector. We show results from a simple Monte Carlo simulation program to indicate how these particles might look in the detector and describe how one might search for these new particles in the ATLAS data stream.

## I. INTRODUCTION

There is a general consensus that the Standard Model (SM) is not complete and that New Physics (NP) is needed to explain some aspects of the SM (e. g., the hierarchy problem and neutrino masses) as well as new data (e. g., dark matter) [1]. In many NP theories (especially GUTS), magnetic monopoles play a role [2, 3]. The motivation for this paper is a model for (composite) elementary particles proposed by Fryberger [4]. This model is based on a non-trivial static solution to Maxwell's symmetric (or generalized) equations, in which magnetic charge and current are explicitly assumed. This particular solution is called a vorton (or quantized vortex) and is described in detail in Ref. [5].

This electromagnetic symmetry of Maxwell's inhomogeneous equations has been called dyality symmetry [6], a name we continue to use here (to avoid confusion with the more common word duality). Dyality symmetry enables a rotation to take place in the generalized electromagnetic charge plane in which the electric strength is along one axis and the magnetic strength is along the other axis. Maxwell's symmetric equations are invariant under this rotation [7-9]. It is a dyality rotation of $\pm\pi/2$ applied to the composite particle model in Ref. [4] that produces a full set of magnetic counterparts to the particles of the SM. The lightest of these with magnetic charge we call a magneticon: a stable spin ½ fermion, which is a counterpart to the electron and whose structure is comprised of a pair of vortons bound electrically.

We also mention here that in order to put magnetic charge on an equal footing with electric charge, Ref. [9] also introduces a magnetic vector potential analogous to the electric vector potential $A_\mu$. This, in turn, implies a second or magnetic photon [10]. This magnetic photon is explored in some detail in Ref. [11]. It is further argued there that the existence of a magnetic photon introduces extra $t$-channel scattering diagrams in the interaction of a magnetic charge (i. e., magneticons) traversing standard electric matter. This additional interaction, in turn, yields an ionization signal that is roughly twice that for minimum ionizing electrically charged particles. This enhanced ionization could also serve as a component of a magneticon signature.

We present below a brief summary of some of the features of the vorton and vorton model as well as an analysis of magneticon production and detection.

*This material is based upon work supported by the U.S. Department of Energy, Office of Science, Office of Basic Energy Sciences, under Contract No. DE-AC02-76SF00515 and HEP.

## II. THE VORTON

The vorton carries an electromagnetic charge of magnitude $Q_v$ and a topological (or Hopf) charge $Q_H = \pm 1$. It has a spherically symmetric charge density distribution $q$ (which is a function of $r$ only) described by

$$q = \frac{4Q_v a^3}{\pi^2 \left(a^2 + r^2\right)^3} \, . \tag{1}$$

The scale of this distribution is characterized by the parameter $a$, the radius of a toroidal coordinate system in Euclidian 3-space. (We use Gaussian units throughout this paper; see Jackson [12].) Note that there are no singularities in this distribution. Like the photon, the vorton has no intrinsic scale; its physical size (that is, $a$) is determined by its creation process in the same manner as the photon creation process determines the photon wavelength. The vorton mass is just that associated with the classical quantity $(\vec{E}^2 + \vec{B}^2)$, which goes like $1/a$.

An important feature of this charge density distribution is that it executes a synchronous double rotation: one rotation is about the $Z$ axis and the other is about a circle (of radius $a$) in the $XY$ plane. The latter rotation results in a toroidal (or smoke ring-like) motion of the charge density. Eq. (1) is invariant under these rotations.

Quantizing (semi-classically) the angular momenta of these rotations to $\pm \hbar$ dictates that $Q_v$ satisfies

$$\frac{Q_v^2}{\hbar c} = 2\pi \sqrt{\frac{3}{5}} \cong 4.867 \, , \tag{2}$$

which is independent of $a$. Thus, the vorton, postulated as a subcomponent of fermions, enables a physical model for point-like fermions. From Eq. (2), the magnitude $Q_v \cong 25.83e$, and it can carry electric and magnetic components, the relative amounts of which are determined by the (sine and cosine of the) dyality angle of the vorton.

## III. THE VORTON PAIR AS FERMION SUBSTRUCTURE

The large intrinsic charge magnitude of the vorton ($25.83e$) and the presence of the dyality angle enable one to consider the magnetically bound vorton pair as a possible model for the known elementary fermions, which notion is pursued in Ref. [4].

First of all, $Q_v$ is large enough to cause a bound vorton pair to collapse to some minimum size (the Planck length?) [13]. And using this concept of a point-like collapsed state, one can model the substructure of the known SM fundamental spin ½ fermions using two magnetically bound vortons in an orbital angular momentum state of $\ell = \frac{1}{2}$.

Fig. 1 illustrates how one can construct the charge of a bare electron (at the Planck scale). The two vortons represented in Fig. 1 have dyality angle values which yield large magnetic charges ($\sim Q_v$) that are equal and opposite (N and S), hence yielding a magnetically neutral sum. Their electric charges are equal, but are of the same sign. Each vorton, then, contributes half the value of the bare electric charge $e_0$, which is larger than $e$, but which would (presumably) be renormalized to the well-known value of $e$ = 4.803204×10⁻¹⁰ esu. The large vorton charge, given by Eq. (2) answers the age-old question asked by Lorentz [14]: What is the force that holds the electric charge of the electron together?

Electron substructure: Two vortons magnetically bound, with dyality angles nearly fully magnetic but each having $e_0/2$

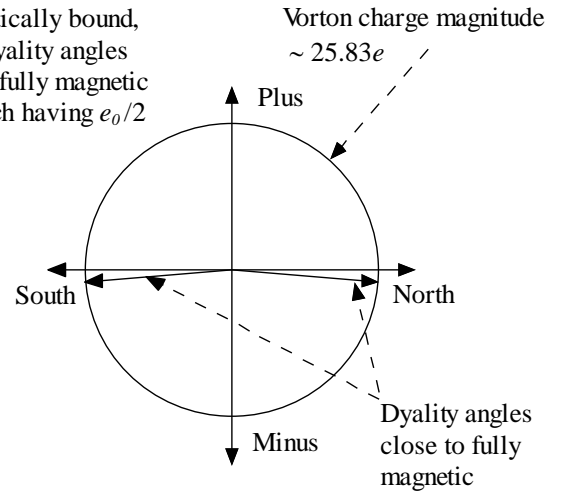

Figure 1: Illustration of how one can construct the electric charge of a bare electron. (This depiction is prior to charge renormalization.)

In Ref. [4] Fryberger shows that there are just enough states to accommodate four generations of spin ½ Dirac four component fermions in the usual isotopic doublet pairings. (In the neutral fermions, the vorton dyality angles would differ by exactly $\pi$.) That is, this model accommodates (predicts?) a fourth generation. And, as mentioned above, by utilizing the dyality symmetry (of generalized Maxwell's equations), this model also enables the description of another complete set of four generations of fundamental spin ½ fermions that have magnetic charge (as well as their neutral isotopic partners). By dyality symmetry, we argue that the charged partners of this set would have a (renormalized) magnetic charge of $1e$.

In this "magnetic" sector, the dyality angles in Fig. 1 would be rotated by $\pm \pi/2$, becoming nearly $\pm$electric; a small deviation from the exact electric axis would lead (in analogous fashion) to the magnetic charges of the bare magnetic fermions. In particular, there should exist a stable magnetic counterpart to the electron, which, as mentioned earlier, we call a magneticon. Of course, one would also expect magnetic muons, magnetic taus, etc. in full analogy to the SM. It is this lightest (stable) charged magnetic fermion and its anti-matter partner that we believe could be copiously pair produced and detected at the LHC. The mass of this magneticon would have to be determined experimentally.

## IV. MAGNETICON PAIR PRODUCTION

In Ref. [11], Fryberger derives that the $e^+e^-$ pair production cross section for magneticon $m\bar{m}$ pairs is

$$\frac{d\sigma_{m\bar{m}}}{d\Omega} = \frac{\alpha\alpha_m(\hbar c)^2 \beta^3}{4s}\left(1+\cos^2\theta\right),  \quad (3)$$

where $\alpha_m = \alpha \cong 1/137$ by dyality symmetry; $\beta$ is the usual relativistic factor in the center of momentum (CM) frame; and $s$ is the CM energy squared.

We assert that such magneticon pair production can occur in all mechanisms that can effect muon pair production (*e. g.,* in $e^+e^-$ colliders through annihilation, and in PP colliders through the Drell-Yan process). And, as one can see from Eq. (3), once energies are well above threshold the magneticon pair production rates will be equal to that of muon pairs. Since muon pairs are being copiously produced at the LHC, we suggest that magneticon pairs are also being copiously produced. The reason they have not been observed as of yet is that no one has explicitly looked for them. (Of course, it is obvious that searches up until now have been strongly influenced by the Dirac prediction that a magnetic monopole would carry a charge (in integral multiples of $68.5e$ [15].)

## V. MONOPOLE SEARCHES

There have been many searches for magnetic monopoles [16-18] since Dirac first postulated that the existence of a magnetic monopole could explain the quantization of electric charge. We have searched the literature to try to find an experiment that might have had a chance of detecting these $1e$ charged monopoles (magneticons). And we have not found any experiments that have looked for magnetic monopoles with ionization level below ~$5e$ for monopole masses above 5 GeV/c².

As mentioned above (and in Ref. [11]) Fryberger has reanalyzed the ionization equations for electric charge and has applied a consistent argument for the ionization signal for a magnetic charge traveling through a standard "electric" medium. Assuming that there is a second or magnetic photon, this analysis indicates that a $1e$ strength magnetically charged particle produces about twice the minimum ionization amount as would a minimum ionizing electrically charged particle. We argue, however, that this additional ionization would not be sufficient to result in the detection of the $1e$ magneticons in the low charge portions of the various searches carried out to date. (We add that unlike electric particles, magnetically charged particles are not expected to have the usual $1/\beta$ ionization dependence at low velocities.)

Consequently, as a general rule, nearly all magnetic monopole searches have required high ionization tracks (>5-6$e$) as part of the monopole signature, which eliminates the possibility of detecting the magneticon as described here. In addition, magneticon tracks do not make "standard" trajectories in detectors with magnetic fields; the magnetically charged particle is accelerated along the direction of the $B$ field and it has no Lorentz $\vec{v} \times \vec{B}$ force term (but rather a $\vec{v} \times \vec{E}$ term). Therefore, magneticon tracks appear to have infinite transverse momentum in solenoidal detector fields. This appearance of high momentum can actually help these events to pass low-level triggers because high momentum muon tracks are generally considered interesting. However, these same tracks in subsequent stages of analysis will not have an acceptable fit to an expected helical track, and hence they will fail the next level of track analysis. OPAL (a LEP and LEP2 detector) did search for magnetic monopoles using track information from their high resolution Jet Chamber [19]. However, the first cut imposed on the candidate events was high ionization (>6$e$), once again eliminating the possibility of detecting the magneticon.

Detectors with no magnetic field should see an excess of muon-like events if the $E_{cm}$ is above the threshold for producing magneticons. The Crystal Ball experiment [20] which ran at the Upsilon (4S) ($E_{cm} = 10.56$ GeV/c²) at DESY did not report any unusual excess of muon events [21]. In addition, CLEO at CESR looked for low ionization magnetically charged (~1-2$e$) tracks and also did not see any events [22]. From these and other low-energy searches we conclude that the magneticon mass is probably above 5 GeV/c².

We have looked extensively at the Free Quark Search (FQS) published data (see Ref. [11]). This was a non-magnetic detector looking for fractionally charged quark signatures. The reason the FQS is interesting to us is that magneticons of low velocity can mimic quarks of low charge. The experiment ran on the PEP-I accelerator, which had an $E_{cm}$ of 29 GeV. It has proven difficult to reach any definite conclusions by looking at their published plots. However, considering the presumed double ionization signature of a magneticon, some mass regions above 5 GeV/c² would seem to be excluded: namely <7.5 GeV/c² and from 11.7 GeV/c² to about 14 GeV/c². That is, the FQS experiment would have a blind spot in magneticon mass in which the low velocity magneticons would fall into their unit electric particle peak, where there were ~13,000 events. A few hundred of these, produced in accordance with Eq. (3), could have been magneticon pairs that would have been missed.

As stated earlier, we have not found an experiment that has conclusively ruled out the existence of the magneticon for masses above 5 GeV/c$^2$ although the FQS experiment appears to disfavor the above mass regions.

## VI. THE ATLAS DETECTOR

Our initial thought was that an $e^+e^-$ collider was needed to detect these new magnetically charged particles. The production rate should be the same as the continuum $\mu^+\mu^-$ production, once it is far enough above the mass threshold so that the $\beta^3$ factor, of Eq. (3), becomes negligible (at unity). However, upon further reflection, it is also clear that the virtual photon in the Drell-Yan process [23] in PP colliders would be a possible production process for these new particles. Again, they would be at much the same rate as muon pair production. With this in mind, we have constructed a simple Monte Carlo generator that uses approximate information about the magnetic detector fields and about the RPC and TGC trigger chambers for the ATLAS detector at the LHC [24, 25]. While we have chosen the ATLAS detector to study, other LHC detectors (especially CMS and LHCb) would also be able to search for these magneticons, as well as would the detectors at RHIC (PHENIX and STAR).

For ATLAS, we note that it is important for these new particles to satisfy the L1 trigger as the ATLAS L1 trigger is very difficult to alter [26]. We have simulated muons and magneticons through the detector solenoid, return yoke field, and the toroidal fields (barrel and endcap) neglecting ionization losses and resolution effects. We use a linear extrapolation between the inner and outer radius magnetic field values of 0.8 and 0.3 T respectively to find the strength of the barrel toroidal field as we trace our particles. In a like manner, we assume that the local endcap toroidal field strength is a linear extrapolation between 1.5 T at the inner radius and 1.0T at the outer radius.

The L1 muon trigger (as we understand it from Ref. [25]) performs an infinite momentum extrapolation from the IP through the hit point of a particular RPC or TGC doublet layer (RPC layer 2 for the barrel and TGC layer 3 for the endcap). Then the difference between the extrapolated point and the track hit point at RPC layer 3 for the barrel and TGC layer 2 for the endcap is taken for high $P_t$ tracks. (Note that we have not included any resolution effects or other factors that might smear these signals.) We produce space points at a nominal radius for the RPC layers or at a nominal $Z$ distance for each TGC layer. The $\eta$ and $\phi$ difference for the extrapolated and trajectory points must be less than a specified number (both less than ±0.1 in Ref. [25]) in order for the L1 trigger to fire. As usual, $\eta$ is the pseudo-rapidity variable, $\eta = -\ln(\tan(\theta/2))$, and $\theta$ and $\phi$ are the standard angular variables in spherical coordinates.

We generate events of either muon or magneticon pairs using a lowest order Drell-Yan function. As a simplification, the generator uses the collinearity approximation (no transverse parton momentum) and the $x1$ and $x2$ values are selected from the plots found in Ref. [23]. These parton distribution plots correspond to a $Q^2$ of $10^4$ GeV$^2$ for both the valence and the sea partons. The muon or magneticon pairs are generated using a $1+\cos^2\theta$ angular distribution in the CM reference frame. We select the $E_{cm}$ range for the production pair to be greater than 50 GeV. If the selected $x1$ and $x2$ values do not meet this requirement we go back and select another pair of values. We use the L1 trigger acceptance mentioned above. The chosen magneticon mass for the histograms below is 7 GeV/c$^2$. We also include a $\beta^3$ rate reduction factor for $E_{cm}$ values near threshold for the magneticons. The $E_{cm}$ of the produced state satisfies the equation $E_{cm}^2 = x1x2s_{beam}$, where $s_{beam} = 4E^2$. $E$ is the beam energy, which we take to be 7 TeV. The state is then appropriately boosted along the $Z$ axis by the energy and momentum of the event.

Figures 2 and 3 show histograms of the generated $E_{cm}$ distribution for 500k events. Fig. 2 is the distribution for muon pair production, and Fig. 3 is for magneticon pair production with a mass of 7 GeV/c$^2$. We stress the simplicity of this generator as we do not include any higher order Drell-Yan terms and use parton distribution functions for only one $Q^2$ value. But we believe this generator is sufficient to obtain a useful comparison with standard muon production in the detector.

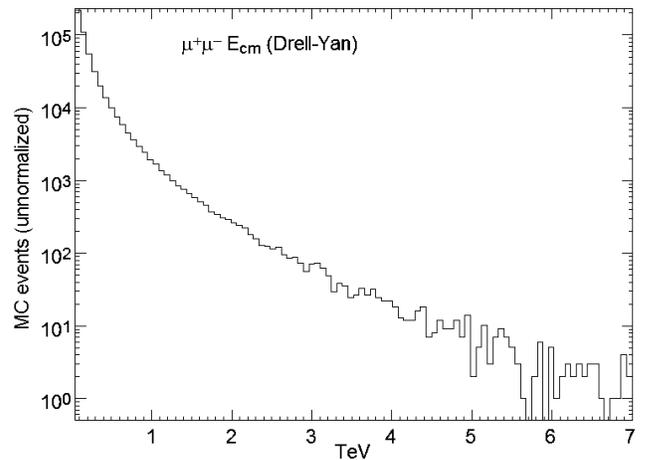

Figure 2: Histogram of $E_{cm}$ for muon pair production using the lowest order Drell-Yan generator mentioned in the text. The LHC beam energy is 7 TeV.

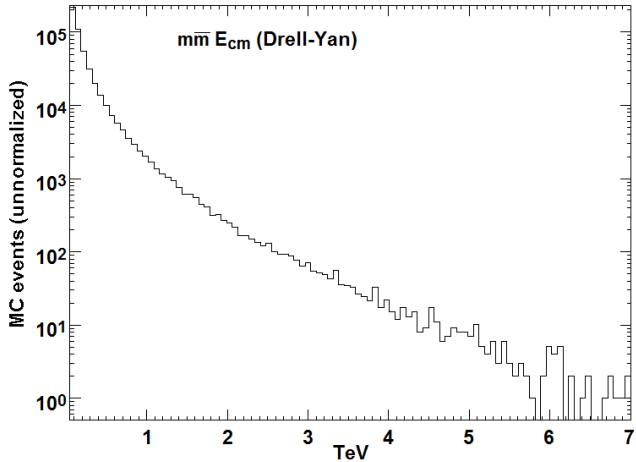

Figure 3: Histogram of $E_{cm}$ for magneticon pair production. The magneticon mass is 7 GeV/c². We used the lowest order Drell-Yan generator mentioned in the text.

Figures 4-7 are histograms of the $\eta$ and $\phi$ difference for the barrel RPC trigger chambers for muon and magneticon events. We generated 500k events each for these distributions. The limits of these histograms are the limits of the L1 trigger acceptance as found in Ref. [25].

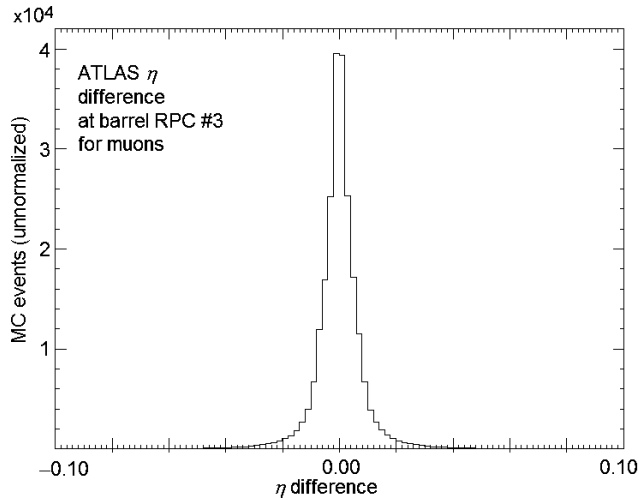

Figure 4: Histogram of the $\eta$ difference in the ATLAS barrel RPC chambers for muons.

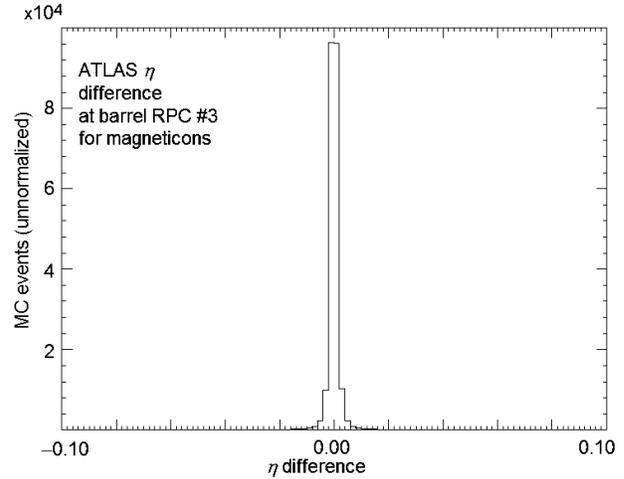

Figure 5: Histogram of the $\eta$ difference in the barrel RPC chambers of ATLAS for magneticons of mass 7 GeV/c².

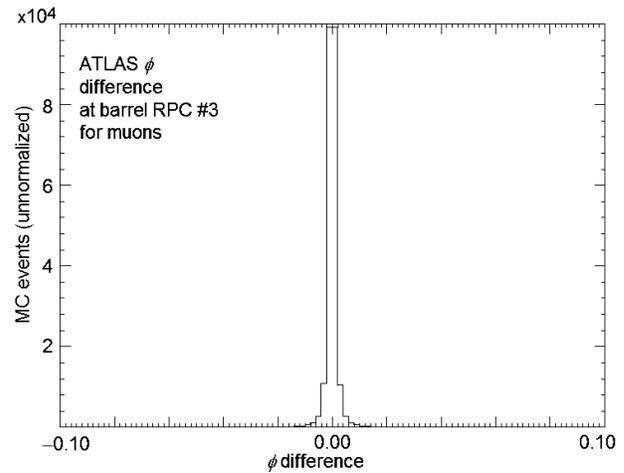

Figure 6: Histogram of the $\phi$ difference in the ATLAS barrel RPC chambers for muons.

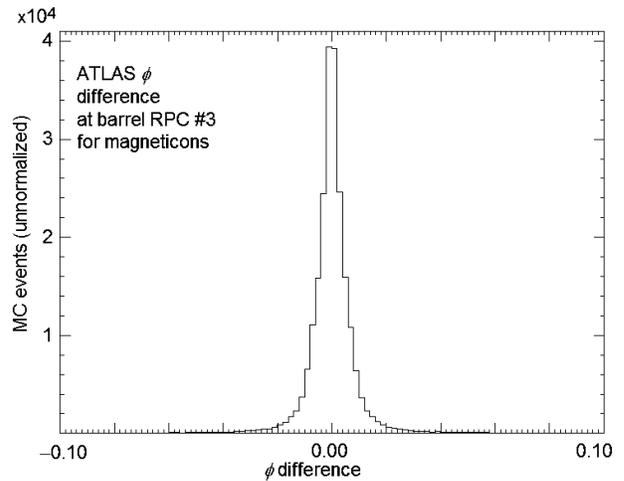

Figure 7: Histogram of the $\phi$ difference in the ATLAS barrel RPC chambers for 7 GeV/c² magneticons.

For the endcap, using the TGC chambers, the same set of histograms for muons and magneticons are even more similar due to the higher momentum track distribution for forward events. It is interesting to note that the $\eta$ distribution for magneticons more closely resembles the $\phi$ distribution for muons and the $\phi$ distribution for muons more closely resembles the $\eta$ distribution for magneticons. In some way, this may reflect the complementary relationship that exists between the electric and magnetic particles. We conclude from these distributions that the L1 trigger of ATLAS should be quite efficient in selecting light mass magneticon events. We estimate an efficiency in excess of 95% of the muon trigger rate for a magneticon mass of 7 GeV/c$^2$.

Table 1 shows the estimated magneticon L1 trigger efficiencies as a function of magneticon mass and $\eta$-$\phi$ difference cuts.

Table 1: Estimated L1 trigger efficiencies for the magneticon normalized to muon trigger efficiencies for various magneticon masses ($m_m$) and for different L1 trigger selections for the ATLAS detector. The values are the ratio (magneticon/muon) events.

| $m_m$ (GeV/c$^2$) | $\eta$ and $\phi$ difference for the L1 trigger | | |
|---|---|---|---|
| | ±0.1 | ±0.05 | ±0.02 |
| 7 | 0.962 | 0.959 | 0.963 |
| 20 | 0.814 | 0.816 | 0.813 |
| 50 | 0.726 | 0.722 | 0.712 |
| 100 | 0.672 | 0.672 | 0.672 |

For Table 1 we restrict our events to be in the barrel. We assume that backgrounds will be lower in the barrel region. For magneticon masses 7 and 20 GeV/c$^2$ the generated $E_{cm}$ lower limit is 50 GeV. For masses 50 and 100 GeV/c$^2$ the $E_{cm}$ limit is 100 and 200 GeV, respectively. The magneticon trigger efficiency in these last two cases will be lower due to additional threshold effects.

One can see from Table 1 that the estimated L1 efficiency for magneticon tracks is quite good even at relatively high magneticon masses.

## VII. MAGNETICON SIGNATURES

### Track Fitting

For the ATLAS detector (or any other detector) a new track fitting algorithm would have to be developed. For ATLAS in particular, events need to be selected that have tracks that have parabolic curvature in the $R$-$Z$ plane when traveling through the solenoidal fields and then have a parabolic curvature in the $\phi$ direction in the toroidal fields. The magneticon tracks are straight in the solenoid end view (they look like they have infinite momentum). We think this feature may be one of the best ways of selecting for these magneticon events. A straight-line track fitter using only the $R$-$\phi$ view information from the solenoid should yield a good selection criterion and be relatively easy to construct. We believe that this straight-line category would not accept too many additional (background) events and thus could be easily added to the general L2 acceptance stream. For the ATLAS detector, the magneticon tracks should also be straight lines in the $R$-$Z$ (or $\eta$) plane where the track goes through the toroidal fields. An event with a track (or tracks) with good straight-line fits in the $R$-$\phi$ plane of the solenoid *and* in the $R$-$Z$ plane of the toroid is surely an excellent magneticon candidate. Only very high-energy cosmic ray muons would look like this. If the constraint that the solenoid track also intersects the collision point is included, we believe that these events should furnish a relatively clean set of magneticon events.

To illustrate how magneticon tracks curve in the $R$-$Z$ view of the solenoidal field we plot low momentum magneticon tracks for the ATLAS detector in Fig. 8. This figure only includes the central solenoid field at 2T and the return flux which we model as a -1T field. This illustration of low momentum tracks is an exaggeration of the magneticon trajectory one might get in ATLAS unless the magneticon mass is very high and the $E_{cm}$ has a very small boost.

Another signature that should be distinctive is the predicted double minimum ionization level for the passage of magneticons through ordinary "electric" material. This may be more difficult to use as an initial event selector but should be helpful as a signature to enhance the analysis of any candidate events.

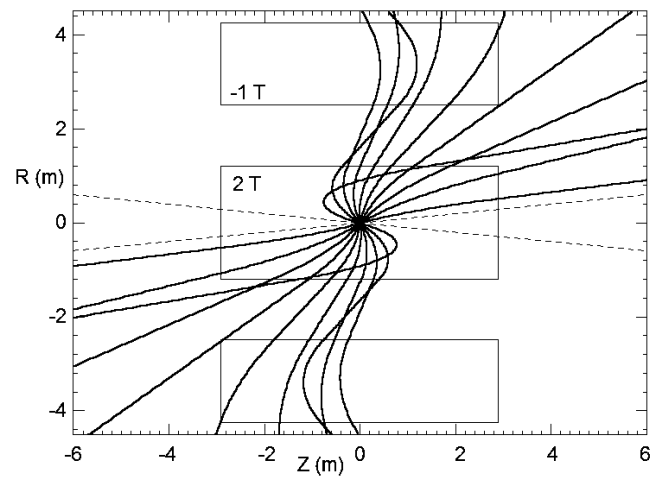

Figure 8: Plot of 10 low momentum magneticon pair events in the $R$-$Z$ plane of ATLAS. In this depiction of the central solenoid, the north magneticons are accelerated to the right and the south magneticons are accelerated to the left. In the return yoke, it is the reverse. We show 10 otherwise equal events that are evenly spaced in $\theta$ with the north magneticon always traveling in the $+R$ direction and the south magneticon in the $-R$ direction as depicted in the Figure. The magneticon mass in this plot is 7 GeV/c$^2$ and the $E_{cm}$ is 15 GeV (close to threshold). This plot illustrates the nature of magnetic monopole tracks in solenoidal fields that one needs to look for in order to search for magneticons. These tracks would appear perfectly straight in the $R$-$\phi$ plane (end view) of the solenoid.

*Monte Carlo Track Simulation*

It is, of course, important to trace magnetically charged particles in the full detector simulation. There may be several ways to do this, and we suggest two possibilities below. The most straightforward procedure would be to add the (presumed) Lorentz force term for magnetic charge in the presence of electric and magnetic fields. That is (in Gaussian units):

$$\vec{F}_m = q_m\left(\vec{B} - \frac{\vec{v}}{c} \times \vec{E}\right). \qquad (4)$$

However, this may prove to be difficult if there are many separate places in the detector simulation program where the forces on a particle are calculated. In addition, one has to introduce magnetically charged particles in the list of particles that can be generated and tracked, and magnetic charge will have to be a new general particle characteristic.

Another possibility is to convert all of the magnetic fields in the detector into electric fields. Then, using an electric force acting on electric particles, with the usual tracking algorithms, one can obtain a reasonably accurate trajectory for magnetic particles using an ordinary electrically charged heavy muon (characterized by the $m_m$ of interest). There would be small perturbative effects from the electric fields already present in any sub-detectors but these field strengths would be very minor compared to the strength of the electric field needed to replace the magnetic field(s). In Gaussian units, the conversion is 1 gauss $\Rightarrow$ 1 statvolt/cm. For SI (mks) units, this relationship becomes 1T $\Rightarrow$ 3×10$^8$ V/m.

## VIII. SUMMARY

We suggest that a search for a low-charge magnetic monopole be performed at all available accelerators. This monopole (called a magneticon) is predicted to be a stable spin ½ fermion with a magnetic charge equivalent to 1$e$. The prediction of the existence of this fermion is the result of explicitly symmetrizing Maxwell's equations with respect to magnetic charge and currents and of assuming a composite electromagnetic substructure for the fundamental fermions of the Standard Model. Based upon an analysis of prior experiments, the mass of this new fermion is expected to be greater than ~5 GeV/c$^2$. Above threshold, it should be produced in the same manner and at roughly the same rates as muon pairs. In particular, the Drell-Yan process in PP colliders should produce these particles as often as muon pairs are produced once one is well above the mass threshold. The ATLAS detector (as well as other detectors) at the LHC presents an excellent opportunity to search for these new particles. We have shown that the unique combination of solenoidal and toroidal magnetic fields in the ATLAS detector allows the formulation of relatively easy and clean search criteria for these magneticons.

## ACKNOWLEDGEMENTS

We would like to thank several people for very helpful discussions and suggestions. Blair Ratcliff and David Muller have suggested several experiments that looked for magnetic monopoles. Su Dong and Rainer Bartoldus have been very helpful in explaining the trigger logic of ATLAS and in helping us to understand how the data stream works. Riccardo Vari has been especially helpful in clarifying the details of the ATLAS L1 trigger for us. David MacFarlane has furnished important references and has been most supportive throughout this effort.

## REFERENCES

[1] J. D. Lykken, "Beyond the Standard Model." *CERN Yellow Report*. pp. 101–109. [arXiv:1005.1676 [hep-ph]]. CERN-2010-002.

[2] G. 't Hooft, "Magnetic Monopoles in Unified Gauge Theories," *Nucl. Phys.* **B79**, pg. 276 (1974).

[3] A. M. Polyakov, "Particle Spectrum in the Quantum Field Theory," *JETP Lett.* **20** pg. 194 (1974).

[4] D. Fryberger, "A Model for the Structure of Point-Like Fermions: Qualitative Features and Physical Description," *Found. Phys.*, **13**, No. 11, 1059 (1983) (SLAC-PUB-2917).

[5] D. Fryberger, "A Semi-Classical Monopole Configuration for Electromagnetism," *Hadronic Journal*, **4**, No. 5, pp. 1844-1888 (1981) (SLAC-PUB-2474).

[6] M. Y. Han and L. C. Biedenharn, *Nuovo Cim.*, **2A**, 544 (1971)

[7] G. Y. Rainich, *Trans. Am. Math. Soc.*, **27**, 106 (1925).

[8] N. Cabibbo, and E. Ferrari, *Nuovo Cim.*, **23**, 1147 (1962).

[9] D. Fryberger, "On Generalized Electromagnetism and Dirac Algebra," *Found. Phys.* **19**, 125 (1989); *Found Phys. Lett.* **3**, 375 (1990).

[10] Jiansu Wei and W. E. Baylis, *Found. Phys. Lett.* **4**, 537 (1991).

[11] D. Fryberger, "On Magneticons and Some Related Matters," in preparation.

[12] J. D. Jackson, *Classical Electromagnetism* (John Wiley & Sons, Inc., New York and London, 1966).

[13] D. Fryberger, *Nuovo Cimento Lett.* **28**, 313 (1980).

[14] H. A. Lorentz, *The Theory of Electrons*, 2$^{nd}$ Ed. (Leipzig, New York, 1915), p. 215.

[15] P. A. M. Dirac, "Quantised Singularities In The Electromagnetic Field," *Proc. Roy. Soc. (London)* **A133**, 60 (1931); *Phys. Rev.* **74**, 817 (1948).

[16] "Review of Particle Physics," *Chinese Physics* **C38**, 1547, ISBN1674-1137, (2014)

[17] V. Dzhordzhadze, "Accelerator Based Magnetic Monopole Search Experiments (Overview)," (2006), http://www0.bnl.gov/npp/docs/pac0307/R20_Physics_Vasily.pdf

[18] K. Milton, "Theoretical and experimental status of magnetic monopoles," *Int. J. Mod. Phys.* **A17**, pg. 732 (2002), [arXiv:0602040v1 [hep-ex]], (2006).

[19] P. Abbiendi, *et al.* "Search for Dirac magnetic monopoles in $e^+ e^-$ collisions with the OPAL detector at LEP2," *Phys. Lett.* **B663**, 37 (2003).

[20] M. Oreglia, et al., "Study of the Reaction $\psi' \rightarrow \gamma\gamma J/\psi$," *Phys. Rev*. **D25**, pg. 2259 (1982).

[21] M. Kobel, *et al.*, "Measurement of the decay of the $\Upsilon(1S)$ and $\Upsilon(2S)$ resonances to muon pairs," *Z. Phys*. **C53**, 193-205 (1992).

[22] T. Gentile, *et al.,* "Search for magnetically charged particles produced in $e^+ e^-$ annihilations at $\sqrt{s} = 10.6$ GeV ," *Phys. Rev.* **D35**, pg. 1081 (1987).

[23] S. Dooling, "Differential Cross Section Measurement of Drell-Yan Production and associated Jets with the CMS," Dissertation, Hamburg University, 2014.

[24] D. Green, Ed., "At the Leading edge: The ATLAS and CMS LHC Experiments," ISBN 13978-981-4304-27-2, 2010.

[25] ATLAS Coll. "Performance of the ATLAS Trigger System in 2010," arXiv:1110.1530v2, (2011), *Eur. Phys. J.* **C72**, 1849 (2012).

[26] D. Su, *Private communication*.